\documentclass{article}
\usepackage{graphicx} 
\usepackage{cite}

\usepackage{amsmath,bm}

\title{Learning optimal erasure of a Static Random Access Memory}
\author{Tomas Basile \& Karel Proesmans}
\date{}

\begin{document}

\maketitle
\section{Abstract}
In this paper, we study the thermodynamic cost associated with erasing a static random access memory. By combining the stochastic thermodynamics framework of electronic circuits with machine learning-based optimization techniques, we show that it is possible to erase an electronic random access memory at arbitrarily fast speed and finite heat dissipation. This disproves a widely held belief that heat dissipation scales linearly with erasure speed.
Furthermore, we find driving protocols that minimize the heat dissipation, leading to explicit design principles for future computer memories. This bridges an important gap between the theoretical framework of stochastic thermodynamics and applications in electronic engineering.

\section{Introduction}
The last couple of decades have seen a dramatic increase in the thermodynamic efficiency of computational devices. More specifically, Koomey's law states that the number of computations, such as erasing a  bit, that one can do with a fixed amount of energy has doubled approximately every $1.6$ years over the last six decades \cite{koomey2010implications}. 

There are however reasons to believe that this scaling will have to come to an end at some point over the next couple of decades. Firstly, Landauer's limit sets a lower bound on the energy necessary to erase one bit of information, which at current rates would be reached by the end of this century \cite{landauer1961irreversibility}. Furthermore, there are a number of reasons to believe that Landauer's limit will never be saturated. Firstly, cache memories, the most energy-consuming memories in modern computers, generally need a constant energy supply, even when not performing any task, which leads to excess heat production. Secondly, a number of finite-time generalizations of Landauer's principle have been derived that suggest that the energy requirements to erase a bit grows linearly with the speed at which the bit erasure is performed \cite{aurell2012refined,proesmans2020finite,zhen2021universal,van2022finite}. If this statement were to be true for real cache memories, it would have profound implications for future computer architectures, as we would reach a fundamental physical limit for standard CMOS technology within the next two decades, where it is physically impossible to further improve computer speed without dissipating more heat. It is therefore important to get quantitative estimates of the fundamental lower limits on the amount of energy needed to operate realistic cache memories, to plan the design and estimate the energy consumption of future CMOS-based computers.

The central goal of this paper is to give such an estimate of one of the most important cache memories, namely a static random access memory cells (SRAM cells). To do this, we will use a model that is both thermodynamically consistent, so that one can apply the framework of stochastic thermodynamics to it, and realistic enough to give quantitative predictions on the fundamental minimal thermodynamic cost associated with operating the memory \cite{freitas2021stochastic,freitas2022reliability}. As our focus is on finding the fundamental physical limits, we will not account for typical engineering problems such as dopant fluctuations and limited control.

By using a recently-developed algorithm that combines results from stochastic thermodynamics with machine learning \cite{engel2023optimal}, we find a protocol that minimizes the amount of heat necessary to erase the memory. We find that the total heat production during the erasure process decreases with increasing erasure speed, disproving a widely held belief in the physics community that heat production should increase with increasing erasure speed \cite{proesmans2020finite,zhen2021universal,ma2022minimal,konopik2023fundamental}. Furthermore, we find that under current design, the minimal amount of heat production is close to the Landauer limit. This suggests that the exponential increase in erasure efficiency could in principle continue for several more decades.

In the next section, we will introduce the theoretical background that is necessary to derive the crucial results. In section \ref{sec:results}, we will subsequently present the main results of this paper, namely the optimal protocols and minimal heat production associated with erasing an SRAM cell. These results are then discussed in Section \ref{sec:discussion}, where we also suggest future research directions.

\section{Theoretical background}
\subsection{Finite-time Landauer principle}
Landauer's principle states that there is a minimal thermodynamic cost associated with the processing of information \cite{landauer1996physical}. More specifically, it states that the amount of heat, $Q$, produced when erasing a bit, i.e., setting the state of the bit that initially has equal probability to be $0$ and $1$ to $1$, is bounded by
\begin{equation}
    Q\geq k_BT\ln 2,
\end{equation}
where $k_B$ is the Boltzmann constant, which we will generally set to $1$, and $T$ is the temperature of the environment. More generally, one has \cite{van2015ensemble}
\begin{equation}
    Q= T\ln 2+T\Delta S,\label{eq:lan2}
\end{equation}
where $\Delta S$ is the total entropy production throughout the process.
Eq.~\eqref{eq:lan2} shows that saturating the bound set by Landauer's principle corresponds to a thermodynamically reversible process, i.e., no net entropy is produced as the entropy increase due to the heat production is exactly compensated by an increase of information on the state of the bit.

Such thermodynamic reversibility can only be achieved for arbitrarily slow processes. Finite-time operation generally comes with excess heat production. This has, over the last decade, led to a number of generalizations of Landauer's principle to finite-time processes \cite{aurell2012refined,proesmans2020finite,zhen2021universal,van2022finite,lee2022speed,rolandi2023finite,proesmans2020optimal,giorgini2023thermodynamic,diana2013finite,zulkowski2014optimal}. For bits whose state can be described by a discrete-state Markov systems, one can write a finite-time generalization of Landauer's principle:
\begin{equation}
    Q\geq T\ln 2+\frac{T}{\left\langle\mu\right\rangle t_f},\label{eq:ftlan}
\end{equation}
where $t_f$ is the duration of the protocol and $\left\langle\mu\right\rangle$ is the expected dynamic activity of the system defined below (cf.~Eq.~\eqref{eq:mudef}). Similar bounds can also be derived for continuous-state systems \cite{aurell2012refined,proesmans2020finite}, quantum mechanical systems \cite{van2022finite} and for erasure with a finite erasure error \cite{proesmans2020optimal}. One common thread with all of these systems is that the heat is bounded by a term inversely proportional to the duration of the protocol, due to an increase in entropy production,
\begin{equation}
    \Delta S\sim t_f^{-1}\Rightarrow Q\sim t_f^{-1},\label{eq:qscale}
\end{equation}
which has also been found in a broad range of toy models and experimental systems, and has been suggested as a universal bound \cite{zulkowski2014optimal,berut2012experimental,jun2014high,diana2013finite,aurell2012refined,proesmans2020finite,zhen2021universal,lee2022speed,boyd2022shortcuts,rolandi2023finite,giorgini2023thermodynamic}.

There are however several reasons to question the universality of this scaling law. It seems, at least mathematically, possible to avoid this scaling by letting $\left\langle \mu \right\rangle$ diverge as $t_f\rightarrow 0$. In fact, there is to the best of our knowledge no fundamental physical law that would forbid such scaling.  Furthermore, measurements in real RAM cells show a linear relation between protocol duration and heat production, rather than an inverse proportionality \cite{hennessy2011computer}. This begs the question of whether these memories do not operate at high enough speed yet to observe scaling of the form of Eq.~\eqref{eq:qscale}, or whether scaling of this form can actually be omitted by increasing the dynamic activity.

\subsection{Thermodynamically consistent model for SRAM cells}
\begin{figure}
    \centering
    \includegraphics[width=\linewidth]{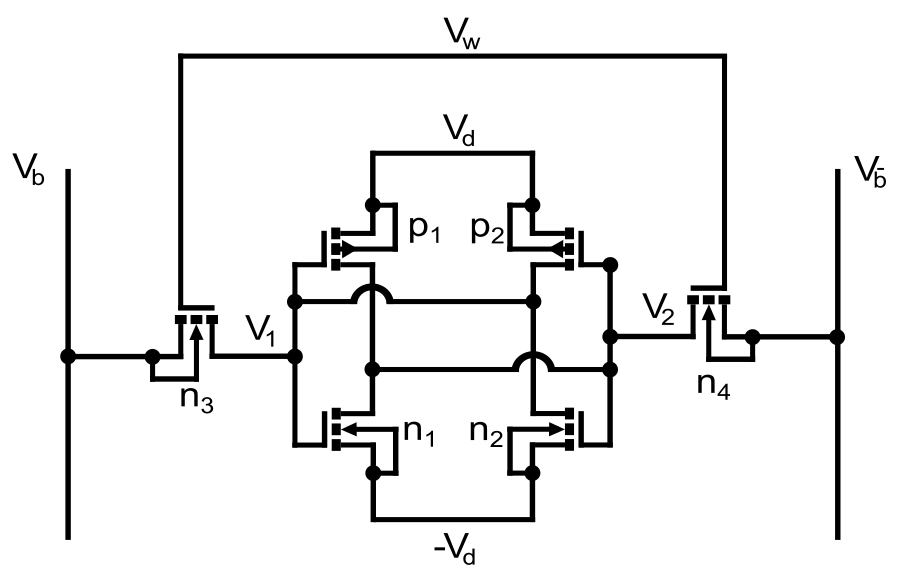}
    \caption{Electronic circuit of a 6t-SRAM memory cell.}
    \label{fig:sram}
\end{figure}

Throughout this paper, we will focus on a 6 transistor SRAM (6t-SRAM) memory cell, which is made up of n-type and p-type transistors connected as shown in Fig~\ref{fig:sram}. The conductors with voltages $V_w$ (the world line), $V_b, V_{\bar{b}}$ (the bit and anti-bit line, where we will always take $V_{\bar{b}} = -V_b$ ) and $\pm V_d$ (driving voltage) are controlled externally, while the conductors with voltages
$V_1, V_2$ may vary and their values are used to define the state of the bit.

To fully understand the dynamics of an SRAM cell, we first need to take a quick look at individual transistors. Transistors have three terminals, known as source $S$, drain $D$ and gate $G$. Electrons move between the source and the drain with a conductivity that depends on the voltage that is being applied at the gate. Particularly, in the sub-threshold regime, the electric current from drain to source for an $n$-type transistor is given by \cite{tsividis2004operation}
\begin{align}
I_n(V_G, V_S, V_D) = \lambda_0 e^{\frac{V_G-V_S}{nV_T}} \left( 1 - e^{\frac{V_S-V_D}{V_T}} \right),\label{eq:Itrans}
\end{align}
while for $p$-type transistors, the electric current is given by $I_p(V_G,V_S,V_D) = -I_n(-V_G,-V_D,-V_S)$. 
Here $\lambda_0$ and $n$ are constants that depend on properties of the transistor, $V_G$, $V_S$ and $V_D$ are the gate, source and drain voltage respectively, and $V_T = T/q_e$ is the thermal voltage, $T$ the temperature of the environment and $q_e$ the elementary charge. From now on, we set the thermal voltage to one, $V_T = 1$, and measure all other voltages with respect to it. Furthermore, we will set $n=1$ to simplify the analysis. Modifying this parameter will not lead to significant qualitative changes to our results.

Using these expressions for current through a transistor, the equations for the expected voltages $V_1$ and $V_2$ of an SRAM cell can be determined using Kirchhoff's circuit law. Moreover, the equations resulting from Kirchhoff's circuit law can be solved analytically in the specific case when 
$V_w \rightarrow -\infty$, so that transistors $n_3$ and $n_4$ have essentially no electrical current flowing and the center part of the circuit is separated from the bit and anti-bit lines. In said case, we get one or three solutions for $(V_1,V_2)$, depending on $V_d$ \cite{freitas2022reliability}:
\begin{align}
V_1 &= V_2 = 0, \\
V_1 &= - V_2 = \pm \ln \left( \dfrac{e^{V_d} + \sqrt{e^{2V_d}-4}}{2}\right) \;\;\;\; (\text{if $V_d \geq \ln 2$}).
\end{align}
The last two solutions only exist if $V_d \geq \ln 2$, and in that case, these two solutions are stable and $V_1=V_2=0$ is unstable.
In the more general case in which $V_w$ is finite, the equations for the circuit cannot be solved analytically, but the conclusions obtained are similar: $V_d$ has to be bigger than a certain value (known as Data Retention Voltage) for the system to have two stable states, one where $V_1-V_2>0$ and one with $V_1-V_2<0$. These two stable solutions can be used to define the two states of the bit, in particular, we say that the bit is in state $0$ if $V_1-V_2>0$ and it is in state $1$ if $V_1-V_2 < 0$.

\begin{figure}
    \centering
    \includegraphics[width=\linewidth]{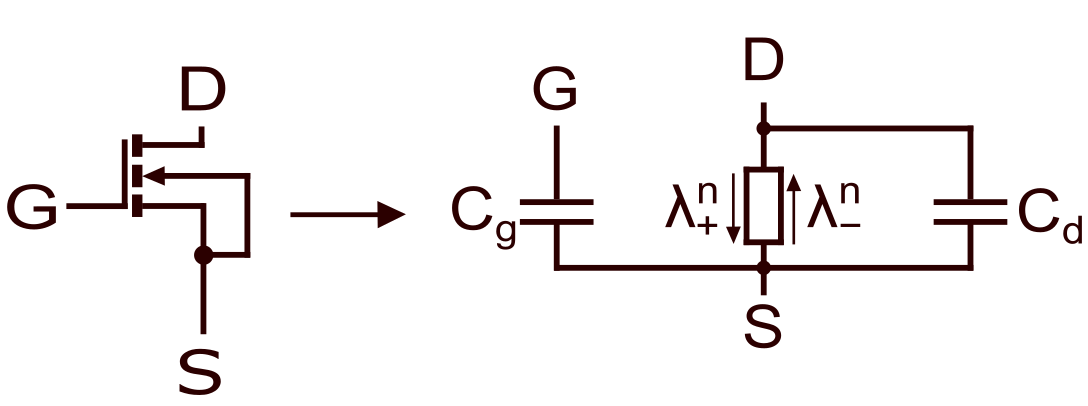}
    \caption{Model of an n-type transistor as a bidirectional Poisson process with parasitic capacitances.}
    \label{fig:transistor}
\end{figure}

If one scales the size of the SRAM cell down to the nanoscale, thermal fluctuations start to become important. This means that the voltages $V_1$ and $V_2$ are no longer deterministic, but should be described by a probability distribution $p(V_1,V_2)$. Furthermore, as the voltage is determined by the number of electrons jumping in and out of the conductors, $V_1$ and $V_2$ can only take discrete values, with differences $V_e$. Here, we will follow the framework introduced in \cite{freitas2021stochastic}. Alternative thermodynamically consistent models for transistors can be found in \cite{gu2019microreversibility,gu2020counting,gao2021principles}. We will model all transistors as in Fig.~\ref{fig:transistor}, that is, we will assume that there are parasitic capacitances between gate and source, $C_g$, and between gate and lead $C_d$. For simplicity, we will ignore other parasitic capacitances and assume that they are the same in each transistor. Under these assumptions, one can use standard methods from electrostatics to show that the voltage change associated with one electron jumping into or out of a conductor is given by $V_e=q_e/(2(C_g+C_d))$.  The evolution of the SRAM cell can now be described by a master equation
\begin{eqnarray}
 \dfrac{d}{dt}p(V_1,V_2)  &= \sum_{\nu,\sigma=\pm}\left(\lambda_\sigma^{(1;\nu)}(V_1-\sigma V_e,V_2) p(V_1-\sigma V_e, V_2)\right.\nonumber\\&\qquad \quad\qquad+ \lambda_{\sigma}^{(2;\nu)}(V_1,V_2-\sigma V_e)  p(V_1, V_2-\sigma V_e)\nonumber
\\&\qquad\left. - \left( \lambda_{\sigma}^{(1;\nu)}(V_1,V_2)  +  \lambda_\sigma^{(2;\nu)}(V_1,V_2)\right) p(V_1,V_2)\right).
\label{eq: master sram}
\end{eqnarray}

Here, the sum $\nu$ runs over all of the different transistors and $\lambda^{(i;\nu)}_{+/-}$ is the rate at which electrons jump in/out of the conductor with voltage $V_i$. These rates need to be chosen, such that Eq.~\eqref{eq:Itrans} is satisfied and need to satisfy local detailed balance for thermodynamic consistency. Detailed calculations and explicit expressions can be found in the Supplemental Information.

Stochastic thermodynamics now predicts that the average entropy production rate is given by
\begin{align}
    \scriptstyle{\left\langle\dot{S}_{\textrm{tot}}\right\rangle=\sum_{V_1,V_2,\nu,\sigma=\pm}\left(\lambda_{\sigma}^{(1;\nu)}(V_1-\sigma V_e,V_2) p(V_1- V_e, V_2)\ln\frac{\lambda_{\sigma}^{(1;\nu)}(V_1-\sigma V_e,V_2) p(V_1- \sigma V_e, V_2)}{\lambda_{-\sigma}^{(1;\nu)}(V_1,V_2) p(V_1, V_2)}\right.}\nonumber\\
   \scriptstyle{ \left.+\lambda_{\sigma}^{(2;\nu)}(V_1,V_2-\sigma V_e) p(V_1, V_2-\sigma V_e)\ln\frac{\lambda_{\sigma}^{(2;\nu)}(V_1,V_2-\sigma V_e) p(V_1, V_2-\sigma V_e)}{\lambda_{-\sigma}^{(2;\nu)}(V_1,V_2) p(V_1, V_2)}\right).}\label{eq:SSRAM}
\end{align}
One can integrate this expression over the duration of the protocol to find the total expected amount of entropy production,
\begin{equation}
    \Delta S=\int^{t_f}_0d\tau\left\langle\dot{S}_{\textrm{tot}}(\tau)\right\rangle.\label{eq:defds}
\end{equation}

On the other hand, the probability for the bit to be in state $1$ is given by
\begin{equation}
    \epsilon=\sum_{V_1,V_2>V_1}p(V_1,V_2).\label{eq:eeSRAM}
\end{equation}
Throughout this paper, we will generally assume that erasing the bit corresponds to setting its state to $0$. In this case, $\epsilon$ corresponds to the erasure error.

The focus of this paper will be on minimizing the energy consumption associated with bit erasure. This corresponds to minimizing the total entropy production, c.f.~Eq.~\eqref{eq:lan2}, while minimising its erasure error, Eq.~\eqref{eq:eeSRAM}. In particular, we will look at protocols that vary $V_w(t)$ and $V_b(t)$ over time to erase the bit. Other control parameters such as $V_d$ will be assumed to be constant throughout the protocol, in agreement with existing computer architectures. To accomplish this minimization, we will apply a machine learning method, based on the algorithm developed in \cite{engel2023optimal}, to the SRAM model. The quantity of interest will be
    $\left\langle \Delta S\right \rangle+\lambda\left\langle \epsilon\right\rangle,$
where $\lambda$ is a Lagrange parameter. Varying it will give minimal values of $\Delta S$ for different values of $\epsilon$. It should be noted that this method only works in regions where the minimal of $\Delta S$ is a convex function of $\epsilon$.
Computational details can be found in the Supplemental Information.

\section{Results\label{sec:results}}
\begin{figure}
    \centering
    \includegraphics[width=\linewidth]{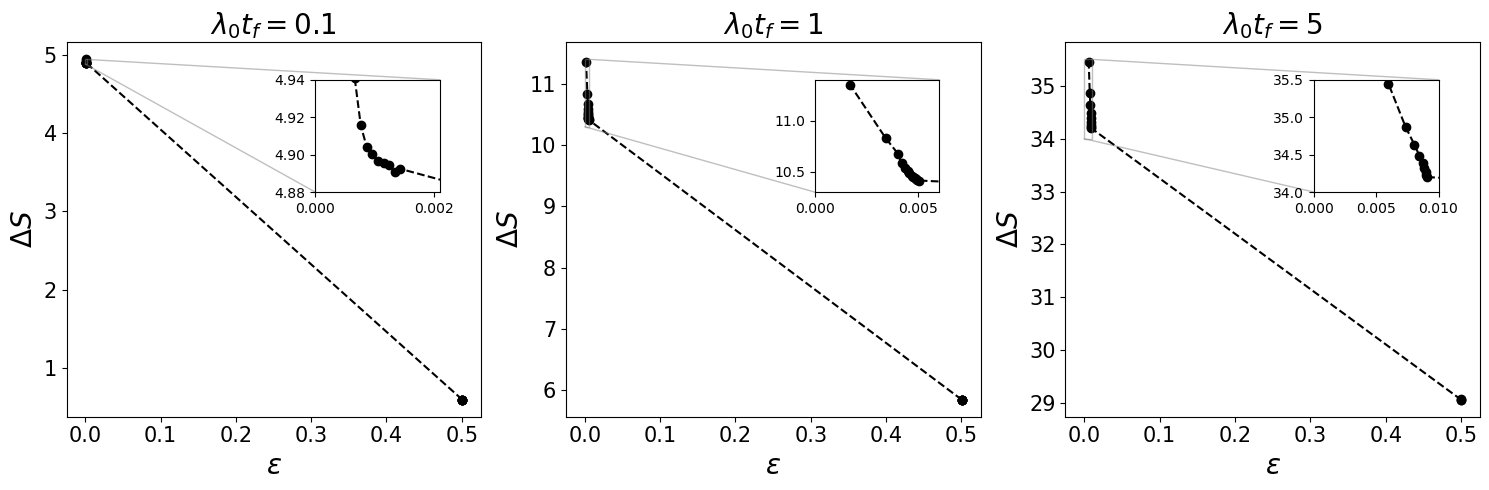}
    \caption{The minimal amount of entropy production, $\Delta S$ as a function of the erasure error $\epsilon$, for $\lambda_0 t_f=0.1,1,5$.}
    \label{fig:epsds}
\end{figure}
Fig.~\ref{fig:epsds} shows the minimal amount of dissipation as a function of the erasure error $\epsilon$ for three different durations of the protocol. One can immediately draw a number of conclusions. Firstly, one can verify that decreasing the erasure error $\epsilon$ generally leads to an increase in the total amount of dissipation, in line with existing results \cite{proesmans2020optimal}. Secondly, we note that the algorithm only manages to capture results either close to $\epsilon=0.5$, corresponding to no erasure, or very close to $\epsilon=0$, corresponding to perfect erasure, due to the non-convexity of the entropy production for intermediate values, $0\ll \epsilon\ll 0.5$. Physically this means that, to minimize the dissipation while keeping $\epsilon$ at an intermediate value, it is more beneficial to stochastically choose between a protocol with very low $\epsilon$ and one with $\epsilon\approx 0.5$, rather than choosing a protocol that always gets the intermediate value of $\epsilon$.

The arguably most important conclusion from the above figures, however, is that $\Delta S$ generally grows with increasing protocol duration. This is in stark contrast with existing models and experiments from stochastic thermodynamics and is a consequence of the volatile nature of an SRAM cell: the SRAM cell will continuously produce entropy, even in the absence of an erasing protocol, leading to a contribution to $\Delta S$ that is extensive in time.

\begin{figure}
    \centering
    \includegraphics[width=\linewidth]{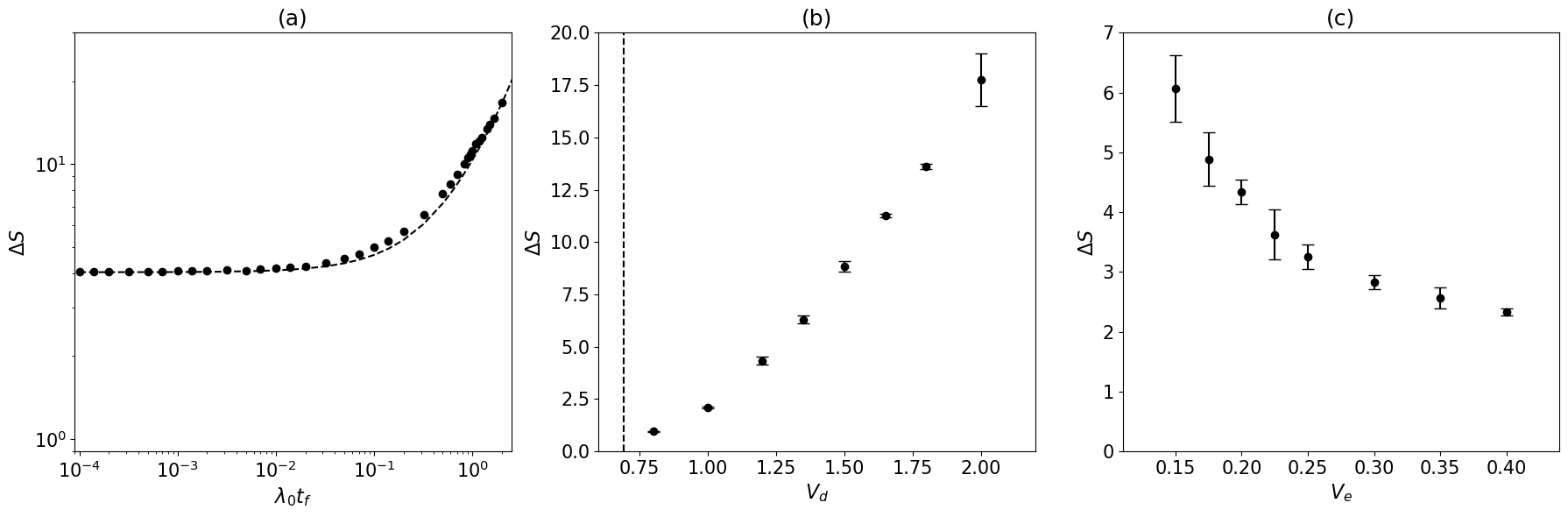}
    \caption{(a) The minimal amount of entropy production, $\Delta S$ as a function of the protocol duration, together with a linear fit, $\Delta S_{\textrm{fit}}(t_f)=a \lambda_0t_f+b$, with $a=6.28 \pm 0.062\, k_B$, $b = 4.04 \pm 0.07\, k_B$, for $\epsilon=0.005$. (b) $\Delta S$ as a function of $V_d$ for $\lambda_0t_f\rightarrow 0$. The dashed vertical line corresponds to the data retention voltage. (c) $\Delta S$ as a function of $V_e$ $\lambda_0t_f\rightarrow 0$.}
    \label{fig:tds}
\end{figure}
This naturally begs the question of how the minimal dissipation to erase a bit depends on protocol duration for our SRAM model. To answer this question, we plot the minimum amount of entropy production as a function of the protocol duration in Fig.~\ref{fig:tds} (a). We find that the entropy production is very well described by a linear function in $t_f$, with a finite offset. Therefore, the conclusion is that speeding up the bit erasure will always lead to a decrease in entropy production in line with the micro-electronics literature. There is however a limit to how much the dissipation can be decreased in this way as the dissipation still seems finite in the limit $t_f\rightarrow 0$.

To further decrease the dissipation, one can try to optimize the system parameters. In particular, we will look at the effect of changing the driving voltage $V_d$ and the parasitic capacitances, which lead to a change in $V_e$. The results are shown in Figs.~\ref{fig:tds} (b) and (c) respectively. Firstly, we see that decreasing $V_d$ generally leads to a decrease in dissipation and that one can decrease it by multiple orders of magnitude by choosing a driving voltage close to the data retention voltage. This does however come at a cost as one can show that the time duration over which the bit can remember its state decreases exponentially with decreasing driving voltage \cite{freitas2022reliability}, suggesting a fundamental trade-off between thermodynamic cost and reliability of the SRAM cell. Furthermore, we see that decreasing the parasitic capacitance, and thereby increasing $V_e$ also reduces the entropy production throughout the erasure. Fig.~\ref{fig:tds} (c) does however suggest that $\Delta S$ levels off at some finite value for increasing $V_e$. From this, we conclude that the total dissipation associated with erasing a reliable SRAM cell is of the order of a few $k_B$ and therefore the heat production associated with it is of the same order of magnitude as the Landauer limit.

\begin{figure}
    \centering
    \includegraphics[width=\linewidth]{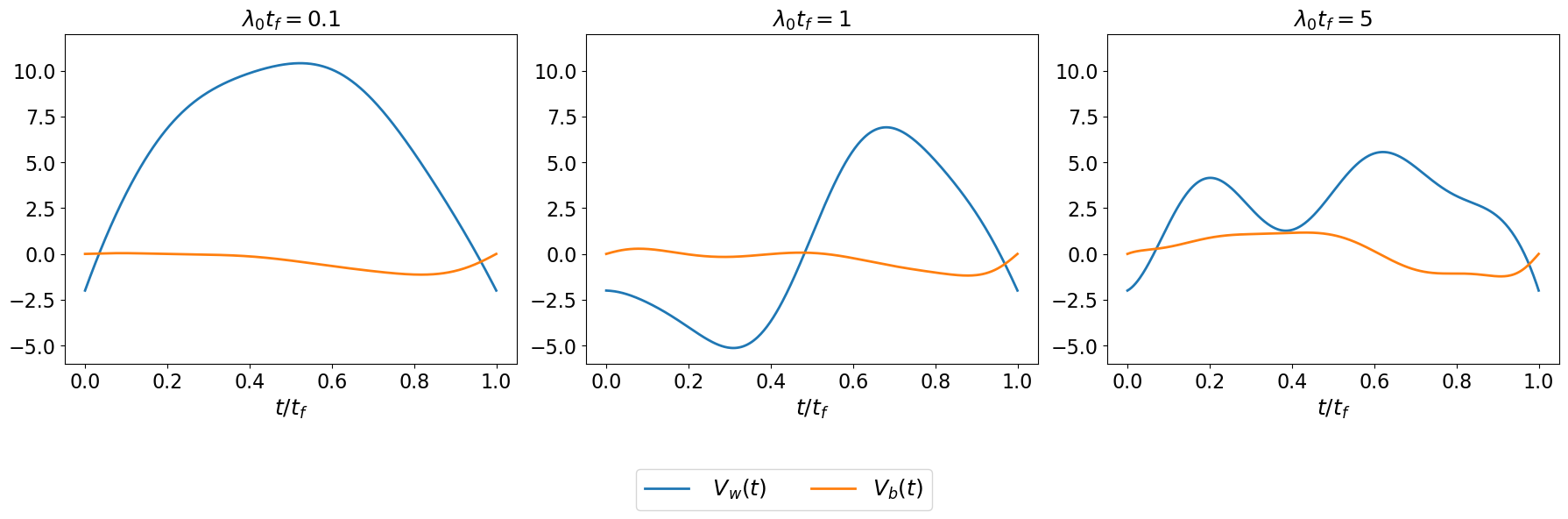}
    \caption{The optimal protocols for $V_w(t)$ and $V_b(t)$ for $\epsilon=0.005$ and $\lambda_0 t_f=0.1,1,5$.}
    \label{fig:optraj}
\end{figure}
We now briefly turn to the optimal protocols for $V_w(t)$ and $V_b(t)$. Fig.~\ref{fig:optraj} shows these protocols for $\lambda_0 t_f=0.1,1,5$. One immediately observes that the voltage over the world line is much larger than that over the bit line and that faster erasure requires larger voltages. Furthermore, we see that the first half of the protocols are very different for the three durations, but that the second half has qualitatively the same behavior: a peak in the voltage over the world line followed by a minimum in the voltage over the bit line. These observations suggest design principles for optimal SRAM erasure.

\begin{figure}
    \centering
    \includegraphics[width=\linewidth]{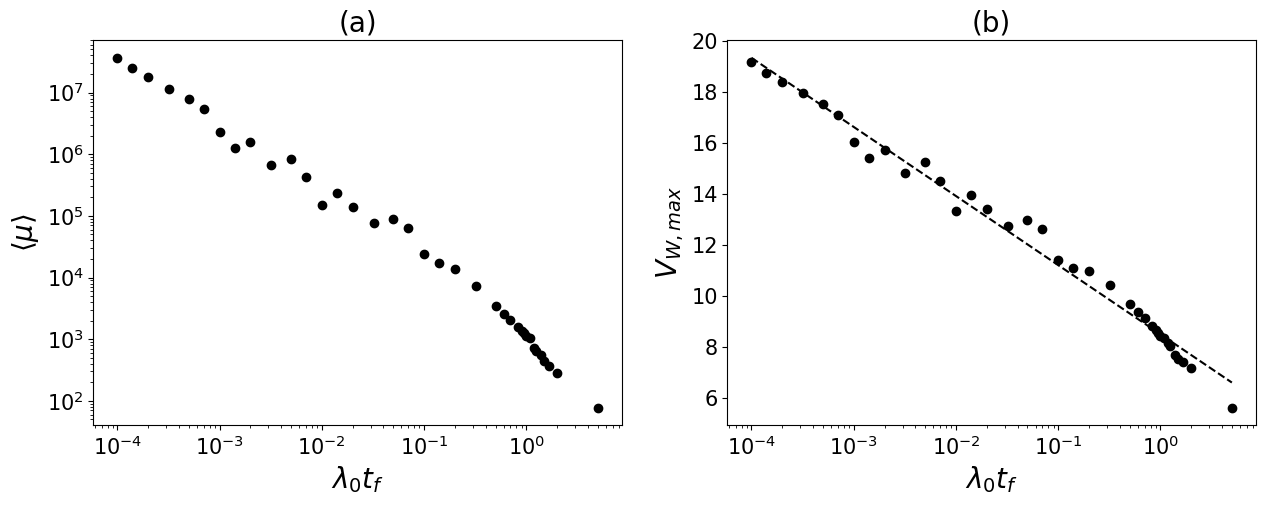}
    \caption{(a) dynamical activity of the optimal protocol as a function of $\lambda_0t_f$. (b) The maximal voltage that the world line reaches as a function of $\lambda_0t_f$. The dashed line corresponds to a logarithmic fit $f(x)=a\ln(x)+b$, with $a=-1.18\pm0.02$ and $b=8.5\pm0.1$.}
    \label{fig:mu}
\end{figure}
Finally, we turn to the question of how these results can be reconciled with the finite-time Landauer principle. Indeed, Eq.~\ref{eq:ftlan} suggests that the entropy production should at least grow inversely proportional to the duration of the protocol in the fast-erasure limit, as long as the dynamical activity,
\begin{equation}
    \left\langle\mu\right\rangle= \frac{1}{t_f}\int^{t_f}_0dt\,\sum_{\nu,V}\lambda_+^{(1;\nu)}(V- V_e,V) p(V- V_e, V)+ \lambda_{-}^{(2;\nu)}(V,V)  p(V, V),\label{eq:mudef}
\end{equation}
is bounded \cite{zhen2021universal}. There is, however, no a priori reason to assume that this dynamic activity is bounded. Indeed, one can calculate the dynamic activity for the optimal protocols for different durations. Results are given in Fig.~\ref{fig:mu}(a). One can verify that for short durations the dynamical activity grows inversely proportional to the duration of the protocol. Looking back at Eq.~\ref{eq:ftlan}, one can then conclude that the finite-time Landauer principle allows for a finite amount of dissipation, even for arbitrary short protocol durations. Although this resolves the main question from a mathematical point of view, one might still wonder how feasible the protocols are from a practical point of view. More specifically, how do the driving voltages scale with decreasing protocol duration. This question is studied in Fig.~\ref{fig:mu}(b). We observe that the maximal necessary driving voltage  of the optimal protocols increases logarithmically when the duration decreases. This relatively minor increase suggests that one can increase the erasure speed a lot without increasing the heat production, by a small increase in driving voltages. For example, for the parameters chosen in Fig.~\ref{fig:mu} at room temperature, to decrease the protocol duration by a factor $10$, one needs to increase $V_W$ by an amount $1.18 V_T\ln 10\approx 0.07\, V$, which is a relatively small increase in driving voltage.

\section{Discussion\label{sec:discussion}}
In conclusion, we numerically studied optimal bit erasure of SRAM cells. Our results bridge the gap between theoretical work in stochastic thermodynamics and practical applications in micro-electronics. Indeed, by using concepts from stochastic thermodynamics, we obtained several practical results regarding realistic bit erasure. 
Firstly, our results show that the energy consumption of SRAM memory cell erasure can in principle be scaled down to values of the order of $k_BT$. This means that there are no fundamental physical reasons to  assume that the exponential increase in computational efficiency, also known as Koomey's law, cannot continue for several more decades.
Secondly, our results show that speeding up bit erasure of SRAM cells leads to a decrease in heat production. This means that there is no trade-off relation between dissipation and speed as suggested in several theoretical papers and toy models. The only cost associated with this increase in speed and efficiency is that the applied voltage needs to be increased logarithmically.
Thirdly, we have derived protocols that minimise the dissipation while erasing the bit in our model. These protocols could serve as qualitative predictions of how to minimise the heat production associated with bit erasure in industrial SRAM cells.
On top of this, our model can be used as a starting point for more advanced models that can give precise quantitative predictions and optimizations of industrial memory devices. To do this, one needs to take several practical limitations into account, such as the limited control over the world and bit lines, dopant fluctuations in the transistors, and drain induced barrier lowering \cite{gul2022sram}.
Furthermore, it would be interesting to look at more advanced SRAM designs, such as 8T or 9T cells, to look at other types of high-speed computer memories such as dynamic RAM (DRAM) cells, or to look at more complex logical designs such as majority-logic decoding.

\bibliographystyle{ieeetr}
\bibliography{main.bib}

\newpage

\section*{Supplemental Information}
\subsection*{Dynamics of SRAM cells}

The transition rates $\lambda_{\pm}^n$ ($\lambda_{\pm}^p$) at which electrons jump from source to drain and from drain to source for a n-type (p-type) transistor depend on the voltages of the gate, source and drain $(V_G, V_S, V_D)$. For these transition rates to correctly describe the electrical current through the transistor, they must produce the current described in Eq.~\eqref{eq:Itrans},
meaning that
\begin{align}
I_n(V_G,V_S,V_D) = q_e (\lambda^+_n (V_G,V_S,V_D) - \lambda_n^-(V_G,V_S,V_D)),
\label{eq: current}
\end{align}
and similarly for a p-type transistor. 

Moreover, the transition rates need to be thermodynamically consistent, meaning that they must satisfy a local detailed balance condition. Assuming that after the electron jumps from source to drain, the voltages change from $V_S$ and $V_D$ to $V_S'$ and $V_D'$, the local detailed balance condition states that
\begin{align}
\dfrac{\lambda_n^+(V_G,V_S,V_D)}{\lambda_n^-(V_G,V_S',V_D')} = e^{\frac{\Delta E(V_G,V_S,V_D)}{k_BT}},
\label{eq: ldb}
\end{align}
and similarly for a p-type transistor. Here $\Delta E(V_G,V_S,V_D)$ is the energetic cost associated with the electron jump. This energetic cost comes firstly from the fact that the electron jumps to a different potential, leading to an energy  change of $q_e(V_D-V_S)$. Secondly, the electron jump leads to a change of the potentials by an amount $V_e = q_e/(2(C_g + C_d))$, which changes the energies stored in the capacitors.  Combining Eqs.~\eqref{eq: current} and~\eqref{eq: ldb} leads to the expressions for the jump rates:
\begin{equation}
\lambda_+^n (V_G, V_S, V_D) = \lambda_0 e^{\frac{V_G-V_S}{V_T}}, \;\;\;\; \lambda_-^n(V_G,V_S,V_D) = \lambda_0 e^{\frac{V_G - V_D - V_e/2}{V_T}}.
\end{equation}
Similarly, the expressions for $p$-type transistors are $\lambda_{\pm}^p(V_G,V_S,V_D) =\lambda_{\pm}^n(-V_G,-V_S,-V_D)$.

Since electron jumps change voltages by an amount $V_e$, if at some moment the state of the cell is given by $(V_1,V_2)$, then the only states it can reach
directly are $(V_1-V_e,V_2)$, $(V_1+V_e,V_2), (V_1, V_2-V_e)$ and $(V_1,V_2+V_e)$.
The first two states correspond to an electron leaving or reaching the conductor of voltage $V_1$ respectively, and similarly for the last two states with the conductor of voltage $V_2$.

We can now find the transition rates between these states. For example, we denote by $\lambda_+^{(1)}$ the rate of the transition that increases $V_1$ by $V_e$. This rate can be easily calculated by considering the different ways in which said increase can happen, that is, the ways in which an electron can move into the conductor with voltage $V_1$. From Fig.~\ref{fig:sram}, we can see that an electron can reach this conductor in three ways, namely, by moving from source to drain of either transistor $n_3$, $p_2$ or $n_2$. Therefore, the transition rate corresponding to an increment of $V_e$ to $V_1$ is given by the sum of the rates of these three transitions:

\begin{align}
\label{eq: lambda+1}
\lambda_+^{(1)}(V_1,V_2) = \lambda_+^p (V_2,V_d,V_1) + \lambda_-^n(V_2,-V_d,V_1) + \lambda_-^n(V_w,V_b,V_1).
\end{align}

The same can be done for the other transitions, so that we get all four transition rates: 

\begin{subequations}
\begin{align}
\label{eq: lambdas 1}
\lambda_+^{(1)}(V_1,V_2) = \lambda_+^p (V_2,V_d,V_1) + \lambda_-^n(V_2,-V_d,V_1) + \lambda_-^n(V_w,V_b,V_1),  \\
\lambda_+^{(2)}(V_1,V_2)  = \lambda_+^p (V_1,V_d,V_2) + \lambda_-^n(V_1,-V_d,V_2) + \lambda_-^n(V_w,V_{\bar{b}},V_2), \\
\lambda_-^{(1)}(V_1,V_2)  = \lambda_-^p (V_2,V_d,V_1) + \lambda_+^n(V_2,-V_d,V_1) + \lambda_+^n(V_w,V_b,V_1), \\
\lambda_-^{(2)}(V_1,V_2)  = \lambda_-^p (V_1,V_d,V_2) + \lambda_+^n(V_1,-V_d,V_2) + \lambda_+^n(V_w,V_{\bar{b}},V_2).
\end{align}
\label{eq: lambdas}
\end{subequations}

\subsection*{Machine-learning algorithm}
The objective of our algorithm is to erase the bit by doing a process of total duration $t_f$. That is, we wish to change the distribution such that at time $t_f$, the probability of being in the bit state $1$, which we denote by $\epsilon$, is as small as possible. To do so, we can control the external voltages $V_w$ and $V_b$ following some protocol $\bm{\lambda}(t) = (V_w(t), V_b(t))$ from $t=0$ to $t = t_f$. With the only condition that the protocol starts and ends at a certain value of $(V_w,V_b)$, which in our case we choose to be $(-2,0)$. 

However, as we not only want to reduce this final probability $\epsilon$, but also do it as efficiently as possible, we also aim to minimize the entropy production $S_i$ over the protocol. To minimize these two quantities simultaneously, we define a loss function that combines them as:
\begin{align}
\label{eq: loss sram}
\omega = \lambda \epsilon + (1-\lambda) S_i,
\end{align}
where $\lambda$ is a number between $0$ and $1$ used to set how much weight we give to entropy production or to the final state of the bit. 

That is, we want to find a protocol $\bm{\lambda}(t) = (V_w(t), V_b(t))$ of duration $t_f$ that starts and ends at $(-2,0)$ and minimizes the loss function $\omega$. We can do so for different values of $\lambda$ to see how the trade-off between minimizing $\Delta S$ or $\epsilon$ changes the result. But for a given value of $\lambda$, how can we find an optimal protocol?

First of all, we need to write a general parametrization of protocols $(V_w(t),V_b(t))$, so as to have parameters to tune in order to optimize the protocol. The parametrization we choose is a linear combination of Chebyshev polynomials:
\begin{align*}
V_w(t) = t(t-t_f) \sum_{i=0}^d \theta_{0i} C_i(2t/t_f - 1) + V_{w0}(t),\\
V_b(t) = t(t-t_f) \sum_{i=0}^d \theta_{1i} C_i(2t/t_f - 1) + V_{b0}(t).
\end{align*}
Here $C_i$ is the ith Chebyshev polynomial, $d$ is the maximum degree of polynomials considered, which in our case was $10$,
and $\bm{\theta}$ are the parameters we wish to optimize. 
$V_{w0}(t)$ and $V_{b0}(t)$ are the initial protocol guesses, which we choose as:
\begin{align*}
V_{w0}(t) &=  -\dfrac{4(V_{w,max}-V_{w,i})}{t_f^2} t^2 + \dfrac{4 (V_{w,max}-V_{w,i}) }{t_f} t + V_{w,i}, \\
V_{b0}(t) &= -\dfrac{4(V_{b,max}-V_{b,i})}{t_f^2} t^2 + \dfrac{4 (V_{b,max}-V_{b,i}) }{t_f} t + V_{b,i}.
\end{align*}
That is, the initial guess $V_{w0}(t)$ makes a parabola starting from $V_{w,i} = -2$, reaching the vertex of $V_{w,max}=8$ at a time $t_f/2$ and returning back to $V_{w,i}$ at time $t_f$. 
Similarly for $V_{b0}(t)$, but with $V_{b,i} = 0$ and $V_{b,max}=-2$. This choice is made to make sure that the bit is erased in the initial guess and therefore the algorithm does not get stuck in a local optimum.
Note that in the parameterization for $V_w(t)$ and $V_b(t)$, the Chebyshev polynomials are evaluated at $2t/t_f -1$ so that the range of values of $t$ are rescaled to $[0,1]$.
Also, the linear combination of Chebyshev polynomials is multiplied by $t(t-t_f)$ to make sure that at $t=0$ and $t=t_f$ the protocol has the same value as the initial guess. 

Then, the algorithm to find the optimal values of $\bm{\theta}$ goes as follows:
\begin{itemize}
    \item[1.] Divide the interval $[0,t_f]$ into $N$ time steps (in our case, we have chosen $N=5\times 10^5$, so that the time steps are of length $\Delta t = t_f / (5\times 10^5)$).
    \item[2.] Set as initial distribution the steady state distribution $p_{ss}(V_1,V_2)$ obtained for the initial values of $V_w,V_b$ (in our case, $V_w=-2$ and $V_b=0$).
    \item[3.] From this starting condition, evolve the distribution $p(V_1,V_2)$  using the master Eq.~\eqref{eq: master sram} with time steps of length $\Delta t$. 
    \item[4.] At each time step  when solving the master equation, calculate the rate of entropy production $\dot{S}$ and add up $\dot{S} dt$ over the whole process to get the total entropy production $S_i$. 
    \item[5.] Having obtained the final distribution $p(V_1,V_2)$, calculate the final probability $\epsilon$ of being in the bit state $1$ by summing over $p(V_1,V_2)$ for the voltages in which $V_1-V_2 < 0$.
    \item[6.] With $\epsilon$ and $S_i$, calculate the loss function $\omega$ as in Eq.~\eqref{eq: loss sram}.
    \item[7.] Through steps $3$ to $6$, use automatic differentiation to obtain the gradient of the loss function, $\nabla \omega$, with respect to the parameters $\bm{\theta}$. Automatic differentiation is a method for calculating gradients of functions at the same time as the function itself is being calculated, by keeping track of the operations made and updating the gradient accordingly.
    \item[8.] Use the gradient to update the parameters $\bm{\theta}$ in the direction of greatest descent. This can be done using various gradient descent algorithms, in particular, we use the Adam algorithm. This completes one step of the gradient descent.
    \item[9.] Continue with the gradient descent steps by repeating steps $3$ to $8$ until reaching some stopping criteria. In this case, we stop after doing $1000$ steps or once we have a difference of $\omega$ of less than one part in a million between steps for $10$ consecutive steps. 
\end{itemize}

\end{document}